\documentclass[fleqn,usenatbib]{mnras}

\usepackage{newtxtext,newtxmath}
\usepackage[T1]{fontenc}
\usepackage{comment}

\DeclareRobustCommand{\VAN}[3]{#2}
\let\VANthebibliography\thebibliography
\def\thebibliography{\DeclareRobustCommand{\VAN}[3]{##3}\VANthebibliography}

\usepackage{graphicx}   % Including figure files
\usepackage{amsmath}    % Advanced maths commands
\title[Variable Polarization Across WR Emission Lines]{
Modeling Variable Linear Polarization Produced by Co-Rotating Interaction Regions (CIRs) Across Optical Recombination
Lines of Wolf-Rayet Stars}

\author[R. Ignace, et al.]{
R. Ignace,$^{1}$\thanks{E-mail: ignace@mail.etsu.edu}
J.E. Bjorkman,$^2$
A.-N. Chen\'e,$^3$
C. Erba,$^{1}$
L. Fabiani,$^{4}$
A.~F.~J. Moffat,$^{4}$
\newauthor R. Sincennes,$^4$
N. St-Louis,$^{4}$ \\
$^{1}$Department of Physics \& Astronomy, East Tennessee State University, Johnson City, TN 37614, USA\\
$^2$ Department of Physics \& Astronomy, University of Toledo, Toledo, OH 43606, USA\\
$^3$ NSF’s NOIRLab/US-ELTP, 670 N. A\'ohoku Place, Hilo, Hawai\'i, 96720, USA \\
$^{4}$D\'epartement de physique, Université de Montr\'eal, C.P. 6128, Succ. C-V, Montr\'eal, QC, H3C 3J7, Canada
}

\date{Accepted XXX. Received YYY; in original form ZZZ}

\pubyear{2023}

\begin{document}
\label{firstpage}
\pagerange{\pageref{firstpage}--\pageref{lastpage}}
\maketitle

\begin{abstract}

Massive star winds are structured both stochastically (``clumps'')
and often coherently (Co-rotation Interaction Regions, or CIRs).
Evidence for CIRs threading the winds of Wolf-Rayet (WR) stars
arises from multiple diagnostics including linear polarimetry.  Some
observations indicate changes in polarization position angle across
optical recombination emission lines from a WR star wind but limited
to blueshifted Doppler velocities.  We explore a model involving a
spherical wind with a single conical CIR stemming from a rotating
star as qualitative proof-of-concept.  To obtain a realistic
distribution of limb polarization and limb darkening across the
pseudo-photosphere formed in the optically thick wind of a WR~star,
we used Monte Carlo radiative transfer (MCRT).  Results are shown
for a parameter study.  For line properties similar to WR~6 (EZ~CMa;
HD~50896), the combination of the MCRT results, a simple model for
the CIR, and the Sobolev approximation for the line formation, we
were able to reproduce variations in both polarization amplitude
and position angle commensurate with observations.  Characterizing
CIRs in WR~winds has added importance for providing stellar rotation
periods since the $v \sin i$ values are unobtainable because the
pseudo-photosphere forms in the wind itself.

%A parameter study involving density contrast for the CIR,
%opening angle, and viewing inclination shows that 
%variability in both polarization degree and position angle 
%at only blueshifted velocities in a snapshot observation can be %produced with the model.  

\end{abstract}


\section{Introduction}
\label{sec:intro}

Precise determinations of massive star mass-loss rates remain elusive, due to time-dependent, non-spherical, non-laminar behavior inferred for
stellar wind flows \citep[e.g.,][]{2008A&ARv..16..209P, 2020Galax...8...60H}.  Uncertainties in the mass-loss rate contribute to ambiguities in our understanding of stellar evolution, since stellar mass-loss quantities are key for understanding angular momentum losses, the evolution of luminosity with age, and stellar remnant outcomes \citep{2014ARA&A..52..487S, 2022arXiv221112153R}.

The wind structure of single, non-magnetic massive stars manifests
itself in two main forms: stochastic and coherent variability.  The
stochastic variability is frequently called ``clumping'' and describes
density variations in the flow that are random
\citep[e.g.,][]{1999ApJ...514..909L, 2008cihw.conf.....H}, 
whereby the wind
may be spherically symmetric in time average, but is formally
aspherical at any given moment. This clumping is often distinguished
as microclumping when clumps are optically thin or macroclumping
or porosity when clumps are optically thick
\citep[e.g.,][]{2007A&A...476.1331O}).  Ultimately, clumps bias
emissive diagnostics used to infer mass-loss rates
\citep[e.g.,][]{1991A&A...247..455H, 2006ApJ...637.1025F,
2013A&A...559A.130S}.  Numerous studies have focused on better
understanding the physics, properties, and statistics of clumps to
infer corrections to obtain more accurate mass-loss rates, $\dot{M}$
\citep[e.g.][]{2014A&A...568A..59S, 2018A&A...619A..59S}.  Such
studies have seen many advances on the topic both observationally
and theoretically \citep[][Brands et al, in prep]{2019MNRAS.490.5921R,
2022A&A...665A..42M, 2023MNRAS.518.5001F}.

The coherent variability is often attributed to Co-rotating Interaction
Regions \citep[CIRs;][]{1984ApJ...283..303M} and have been associated with the ubiquitous Discrete Absorption Components (DACs) observed in the UV P Cygni line absorption components of strong resonance lines of O~stars \citep[e.g.,][]{1989ApJS...69..527H, 1996A&AS..116..257K}.  The CIRs are believed to be associated with starspots that produce local differentials in mass flux and in the wind flow speed
profile (i.e., ``velocity law'') at the wind base \citep[e.g.,][]{1996ApJ...462..469C, 2004A&A...423..693D, 2017MNRAS.470.3672D}. Shocks form with hypersonic wind speeds, leading generally to spiral features -- the CIRs --
that thread the flow \citep{2015ApJ...809...12M}.  As with clumping, CIRs produce time-dependent behavior and bias emissive diagnostics.  Unlike clumping, these effects
are phase dependent and cyclic, \citep{2004A&A...413..959B}, with slow evolution as the starspots evolve.  Since the identification of CIRs, there have been numerous studies focused on their origins and observable consequences, with several recent advances in terms of modeling effects for multi-wavelength diagnostics \citep{2019MNRAS.489.2873C, 2020MNRAS.497.1127I}.

Clumping is universal to all hot-star winds, while CIRs are universal to O-star winds.
Evidence for CIRs in the dense winds of the massive Wolf-Rayet (WR) stars has been amassing for years \citep{2011ApJ...735...34C, 2011ApJ...736..140C, 2016MNRAS.460.3407A, 2018MNRAS.474.1886S}.  The winds are so dense that 
the inner hydrostatic layers are unobservable. 
In particular, direct information from spectral lines about rotational speeds such as $v \sin i$
values are simply unobtainable. Yet, elucidating the properties of rotation for WR stars is crucial for understanding
supernovae, gamma-ray bursts, and black hole remnants \citep{2005A&A...443..643Y, 2009ARA&A..47...63S, 2012A&A...542A..29G, 2017A&A...603A.120V}.  Confirming the presence of a CIR in any given WR star, and measuring its cyclic behavior
provides valuable information about the stellar rotation period, since the whole CIR feature revolves at constant angular speed.

A powerful tool for probing the geometry of unresolved stellar sources is polarimetry \citep[e.g.,][]{2012AIPC.1429.....H}.  Specifically, intrinsic linear polarization can only arise if the source deviates from spherical.  For hot massive stars, electron scattering is a key polarigenic opacity, both in the atmospheres and the winds.  However, polarization observations are complicated by a wavelength-dependent contribution from the interstellar medium, often modeled with the ``Serkowski Law''  \citep{1975ApJ...196..261S}.
For a star with
an intrinsic continuum polarization, it is well-known that the polarization is depressed across emission lines when the line photons are little scattered by the copious free electrons \citep{1979ApJ...231L.141M}.  This effect of 
depressing the polarization is known as the ``line effect'' and can aid with distinguishing between the intrinsic stellar polarization and the interstellar contribution \citep{1991ApJ...382..301S}.  
The intrinsic polarization position angle (PA) of the stellar continuum is generally different from that of the interstellar polarization.  In particular, the PA from the interstellar contribution is fixed at all wavelengths.  
Consequently, if the PA is observed to change (or ``rotate'') across the line, the effect certainly is intrinsic to the star.  

\begin{figure}
    \includegraphics[width=\columnwidth]{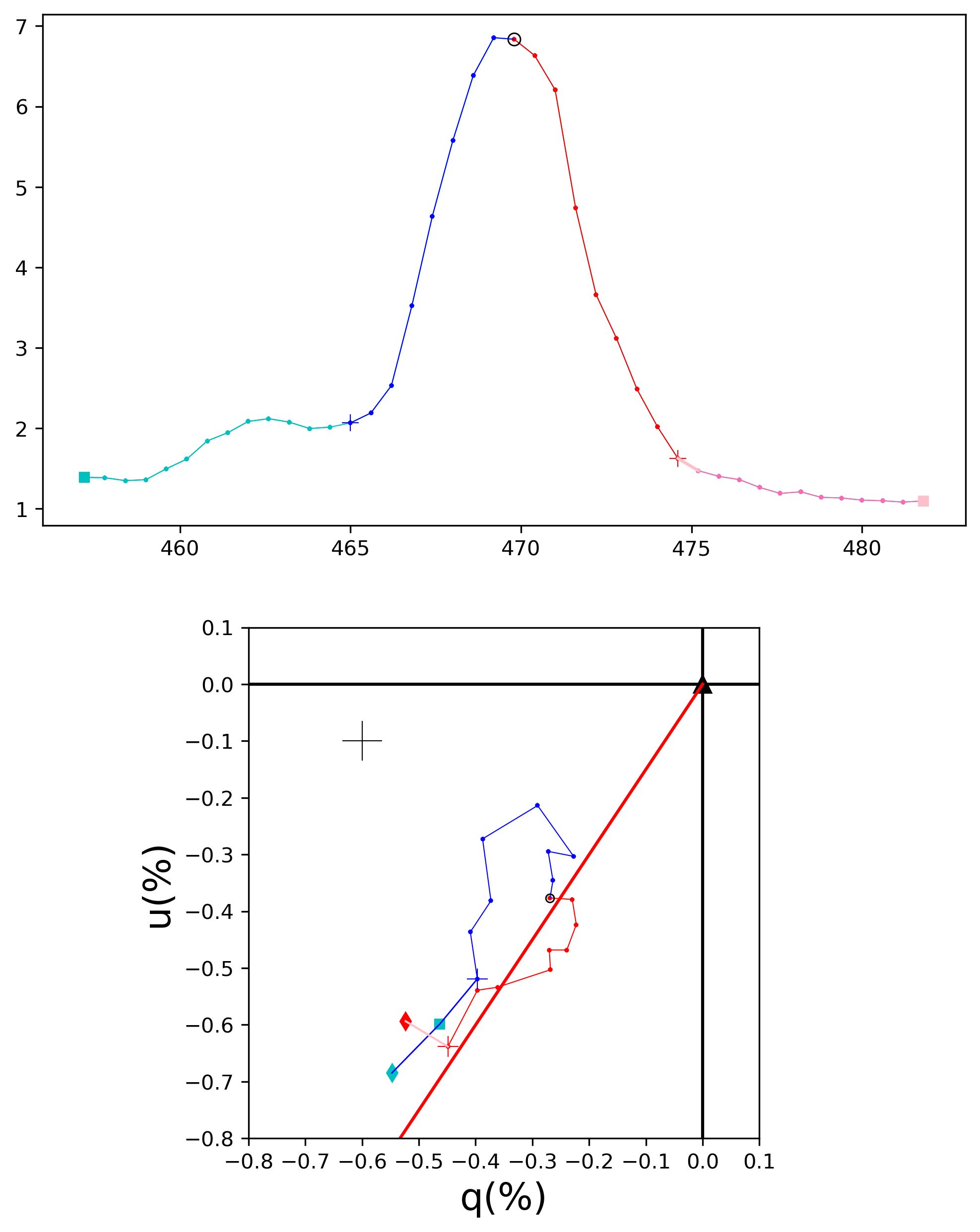}
    \caption{Based on the observational data of WR~6 for He~{\sc ii}~4686 from \citet{1990ApJ...365L..19S}, the upper panel shows the emission line profile while the lower panel shows polarization across the line.  At top: The circle near line peak is the line center.  Color indicates blue- and redshifted sides.  At bottom: A plot of $q-u$ variations across the line, binned to obtain a typical error as indicated toward upper left.  The three diamonds (1 red and 2 blue) are for continuum polarizations.  The two large crosses identify the largest blue- and redshifted velocities attributed to the line.  The straight red line is drawn to start from the origin and roughly follow the redshifted velocity points.  Note that the prominent loop is at blueshifted velocities, whereas polarimetric variations across the red side follow mainly a straight line (see text).}
    \label{fig1}
\end{figure}

As an example, 
in Figure~\ref{fig1} we present an adaptation\footnote{Archival data obtained from the Mikulski Archive for Space Telescopes (MAST) at archive.stsci.edu/hpol.} of Figure~2 from \cite{1991ApJ...382..301S} for WR~6 showing the He~{\sc ii} 4686 recombination line shown in the upper panel and for polarization across the line in the lower panel. As concluded by the authors, a loop is detected in the $q-u$ plane that appears primarily on the blueshifted side of the line, in other words the PA rotation is limited to blueshifted velocities for a given epoch.  By contrast polarization for the redshifted points in the line are largely at fixed PA.  Similar effects are observed in more recent data obtained with ESPaDOnS (Fabiani et al, in prep).

For an expanding symmetric wind, it is only the flow in front of
the star that contributes to the absorption profile, and that
absorption only appears at blueshifted velocities. This has motivated
us to present a simplified wind model with a conical CIR that rotates
with the star. Its phase-dependent projection against the stellar
pseudo-photosphere, along with its density contrast compared to the rest
of the wind, produces differential absorption of the limb polarization
of the pseudo-photosphere. We develop this model and conduct
a parameter study as proof-of-concept for producing PA rotation at
blueshifted velocities along with polarimetric variation at redshifted
velocities for constant PA.

The paper is structured as follows.  Our model assumptions are
introduced in section~\ref{sec:model}, including the assumptions
of the wind and the CIR plus a review of the Sobolev approximation
for the radiative transfer of emission lines as applied to the case
at hand.  This is followed by a parameter study presented in
section~\ref{sec:study}.  Line profile variability along with
polarization variability -- both degree of polarization and
polarization position angle -- are simulated as functions of viewing
inclination and properties of the CIR.  A summary is given in
section~\ref{sec:summary}.  Finally, the details of the Monte Carlo
radiative transfer calculations to ascertain realistic limb
polarization and limb darkening distributions for pseudo-photospheres
of thick scattering winds are presented in an Appendix.

\section{Line Profile Modeling}
\label{sec:model}

This paper is inspired by suggestions that there is a position angle (PA) rotation across the line in linear polarization at blueshifted velocities of an optical recombination emission line, but a lack of related behavior at redshifted velocities.  When polarimetric variability spans the full range of Doppler shifts for circumstellar bulk flows, one typically attributes such effects to asymmetric structures in the scattering volume \citep[e.g.,][]{1977A&A....57..141B, 1978A&A....68..415B, 2000A&A...356..619B}.  By contrast, for an expanding outflow, the most natural culprit for PA rotation limited to blueshifted velocities in the line but not redshifted ones would be absorption effects.  While acknowledging that WR winds are complicated in terms of having CIRs (perhaps more than one), a stochastic component in terms of clumps, the formation of wind shocks to account for X-ray emission, and optically thick radiatve transfer in continuum opacity (i.e., electron scattering and free-free/bound-free), here we develop a limited model in terms of a single CIR that co-rotates with the star to produce polarimetric variability at blueshifted velocities but not redshifted ones.  This is achieved through time-dependent, but cyclical, differential absorption of the limb polarized pseudo-photosphere by a CIR structure that is overdense compared with the rest of the wind.

\subsection{Coordinate System Definition}

We introduce a system of spherical coordinates $(r, \chi, \psi)$ from the observer's perspective centered on the star, where
$\chi=0^\circ$ is in the direction of the observer and defines the
$z$-axis. The star is taken to have spherical coordinates
$(r,\theta,\phi)$, with $\theta$ as the colatitude defined from the spin axis of the star $z_\ast$. The viewing inclination $i_0$ of the
star is expressed in terms of unit vectors as $\hat{z}\cdot\hat{z}_\ast = \cos i_0$.

To explore the effects of a CIR in producing loops in the Q--U plane across only blueshifted velocities for an expanding wind, we consider
stellar wind and CIR models that are simplified to demonstrate a proof-of-concept result and to conduct a parameter study. We assume the stellar wind is spherically symmetric outside the CIR, and that it expands according to a linear velocity law, with

\begin{equation}
\label{eq:linearvlaw}
v(r) = v_{\rm phot}\times\left(\frac{r}{R_{\rm phot}}\right),
\end{equation}

\noindent where $R_{\rm phot}$ is the radius of the 
pseudo-photosphere formed in the wind itself, taken to be where
optical depth unity is achieved along the observer's line-of-sight
(LOS) to the star.  For typical WR winds, we expect $R_{\rm
phot}\sim 2 R_\ast$, for $R_\ast$ the hydrostatic radius of the
star.  At this location, the wind speed at the photosphere, $v_{\rm
phot}$, will be a few hundreds of km/s.

For the CIR we assume a strictly conical structure with
a density deviation from the spherical wind by a factor of $1+\eta$ \citep{2015A&A...575A.129I},
where $\eta$ is a dimensionless quantity with $\eta \ge -1$.
When $\eta$ is negative, the CIR density is lower than the density of the 
wind; $\eta=0$ is equivalent to no CIR; and positive $\eta$ 
is for a CIR that is overdense compared to the wind. The density of the CIR is thus given by

\begin{equation}
\rho_{\rm CIR} = (1+\eta)\,\rho_{\rm sph},
\end{equation}

\noindent where the spherical wind density is

\begin{equation}
\rho_{\rm sph} = \frac{\dot{M}}{4\pi\,r^2\,v(r)} = \rho_{\rm phot}\,
	\xi^3.
\end{equation} 

\noindent For the wind density, the linear velocity law from equation~(\ref{eq:linearvlaw}) was used, and a normalized inverse radius  introduced as 

\begin{equation}
\xi = R_{\rm phot}/r.
\end{equation}

\noindent The expansion velocity within the CIR is taken as the same as external to the CIR.  Consequently, our model takes the wind expansion to be spherically symmetric everywhere, but the density distribution is asymmetric via the geometry of the CIR through the factor $\eta$.

\subsection{The Sobolev Approximation for Recombination Emission Lines}

To simulate line profiles for heuristic purposes, the Sobolev approximation is adopted \citep{1960mes..book.....S}.
The line profile shape is calculated by evaluating optical depths
on isovelocity surfaces.  These surfaces represent the locus of
points for which the Doppler shift in the direction of the observer is constant. The line-of-sight (LOS) velocity shift is given by

\begin{equation}
v_{\rm z} = -v(r)\,\hat{z}\cdot \hat{r} = -v(r)\,\cos \chi
	= -v_{\rm phot}\times \left(\frac{z}{R_{\rm phot}}\right).
\end{equation}

\noindent This equation indicates that surfaces with constant $v_{\rm z}$ correspond to planes oriented with constant $z$, thus normal
to $\hat{z}$. We introduce a normalized velocity shift as

\begin{equation}
w_{\rm z} = v_{\rm z}/v_{\rm phot}.
\end{equation}

\noindent In normalized units, the
isovelocity surfaces intersect the photosphere for $-1 \le w_{\rm z}
\le +1$.  Note that unlike typical wind velocity laws, a wind with homologous
expansion formally has no terminal speed.  

For the radiative transfer, we introduce the Sobolev line optical depth
with

\begin{equation}
\tau_L = \frac{\kappa_L\,\rho\,\lambda_L}{|dv_{\rm z}/dz|},
\end{equation}

\noindent where $\kappa_L$ is the frequency-integrated opacity
with units of ${\rm cm~g^{-1}~Hz}$, $\rho$ is the density, $\lambda_L$
is the rest wavelength of the line of interest, and $dv_{\rm z}/dz$
is the LOS velocity gradient.   Seeking to model a recombination
line that forms in the wind, $\tau_L$ scales with the square
of density.  However, NLTE effects and a radius-dependent temperature
can introduce additional radius dependencies.  Consequently, we use
a scaling relation of

\begin{equation}
    \kappa_L\,\rho \propto \rho^2(r) \, g(r),
    \label{eq:kappa}
\end{equation}

\noindent where $g(r)$ can be used as a fitting parameter for the
line profile shape as needed.  

For a linear velocity law, the LOS velocity gradient simplifies to \citep[e.g.,][]{1978stat.book.....M}

\begin{equation}
\frac{dv_{\rm z}}{dz} = \frac{v(r)}{r}\,\sin^2\chi
	+ \frac{dv}{dr}\,\cos^2 \chi = \frac{v_{\rm phot}}{R_{\rm phot}}.
\end{equation}

\noindent As a result, the optical depth becomes

\begin{equation}
\tau_L = \tau_0\,\xi^6\,g^2(\xi)\,\left[1+\eta(r,\chi,\psi)\right]^2.
\end{equation}

\noindent where $\eta=0$ when not in the CIR, and $\eta \ne 0$ when
inside the CIR.  Also, $\tau_0$ is a line optical scale
used as a free parameter of the model.

We model the effect of a CIR for the line transfer entirely in terms
of density contrast $\eta$.  However, hydrodynamic models
\citep[e.g.,][]{1996ApJ...462..469C} indicate that the transition
in density between a CIR and the wind is not discontinuous.  Moreover,
the velocity field interior to the CIR does not follow that of the
wind and can be non-monotonic with large velocity gradients.  These
gradients can be important in modeling variable UV P Cygni lines.
Despite these shortcomings in our approach, we are modeling
recombination emission lines from WR~winds that generally
display no net P Cygni absorption \citep[unlike scattering
resonance lines,][]{1995A&AS..113..459H}, plus our intent is mainly
proof-of-concept.  Our model explores the asymmetric and time-dependent
changes in the Sobolev optical depth for producing Q--U trends
across recombination lines.  We seek to demonstrate the effect and
conduct a parameter study in order to inform future, more detailed
model calculations.

\subsection{Emission Line Profile Shape}

Calculation of the line profile shape is determined by
the intensity reaching the observer.  We integrate over intensity
to formulate the line luminosity as a function of velocity shift.
For a distant observer, we evaluate intensities on a grid of
rays parallel to the observer $z$-axis.  We introduce a cylindrical
radius $\varpi$ as the impact parameter of a ray and normalized to
$R_{\rm phot}$.  The radiative transfer for the emergent monochromatic intensity becomes

\begin{equation}
I(w_{\rm z},\varpi,\psi) = I_C\,e^{-\tau_L} + S_L\,\left( 1 - 
	e^{-\tau_L}\right),
	\label{eq:intensity}
\end{equation}

\noindent where the subscript $\nu$ denoting frequency dependence has been omitted, $I_C$ is the stellar continuum intensity for those rays that intersect the
photosphere, $S_L$ is the line source function, and $\tau_L$ is
the Sobolev line optical depth.  Rays only intercept the photosphere
when $\varpi \le 1$.  For rays with $\varpi > 1$ do not
intercept the photosphere, and the first term can be ignored.
Additionally, occultation of material behind the star is not included in the flux calcuations for rays intercepting the photosphere.

For a spherical wind, the emergent intensity is a function only of
$\varpi$ (by symmetry) and of $w_{\rm z}$, by virtue of the velocity
shift and corresponding isovelocity surface under consideration.
Here, however, the intensity also depends on $\psi$ because of the CIR.
This means, for example, that an intensity or isophotal map will not be centro-symmetric about the observer LOS.

In the next section, our method for calculating the linear polarization
across the line in terms of Stokes parameters will be introduced.
Here, we note that the emergent intensity in the preceding equation
is for Stokes~$I$.  In this Stokes parameter, the luminosity of
line emission is

\begin{equation}
\frac{L_I(w_{\rm z})}{L_\ast} = \int\,e^{-\tau_L}\,\varpi\,d\varpi\,\frac{d\psi}
	{\pi} + \int \frac{S_L}{I_C}\,\left( 1 - e^{-\tau_L}\right)\,
        \varpi\,d\varpi\,\frac{d\psi}{\pi},
            \label{eq:LI}
\end{equation}

\noindent where again the first term is for the absorption of the
photosphere for rays that intercept it, and the second term is for the emission produced in the wind itself.  Note that the integrations are
carried out over the isovelocity surfaces.

In the core-halo approach, the stellar luminosity $L_\ast$ is expressed as

\begin{equation}
L_\ast = 8\pi^2\,R^2_{\rm phot}\,\int\,I_C(\varpi)\,\varpi\,d\varpi.
\end{equation}

\noindent Our prescription for
$I_C(\varpi)\equiv I_\ast\,F(\varpi)$, which includes limb darkening, is described in the Appendix, where $I_\ast$ is a fiducial intensity scale and $F$ is the intensity distribution across the pseudo-photosphere formed in the thick scattering wind. 

\begin{figure}
\includegraphics[width=3.5in]{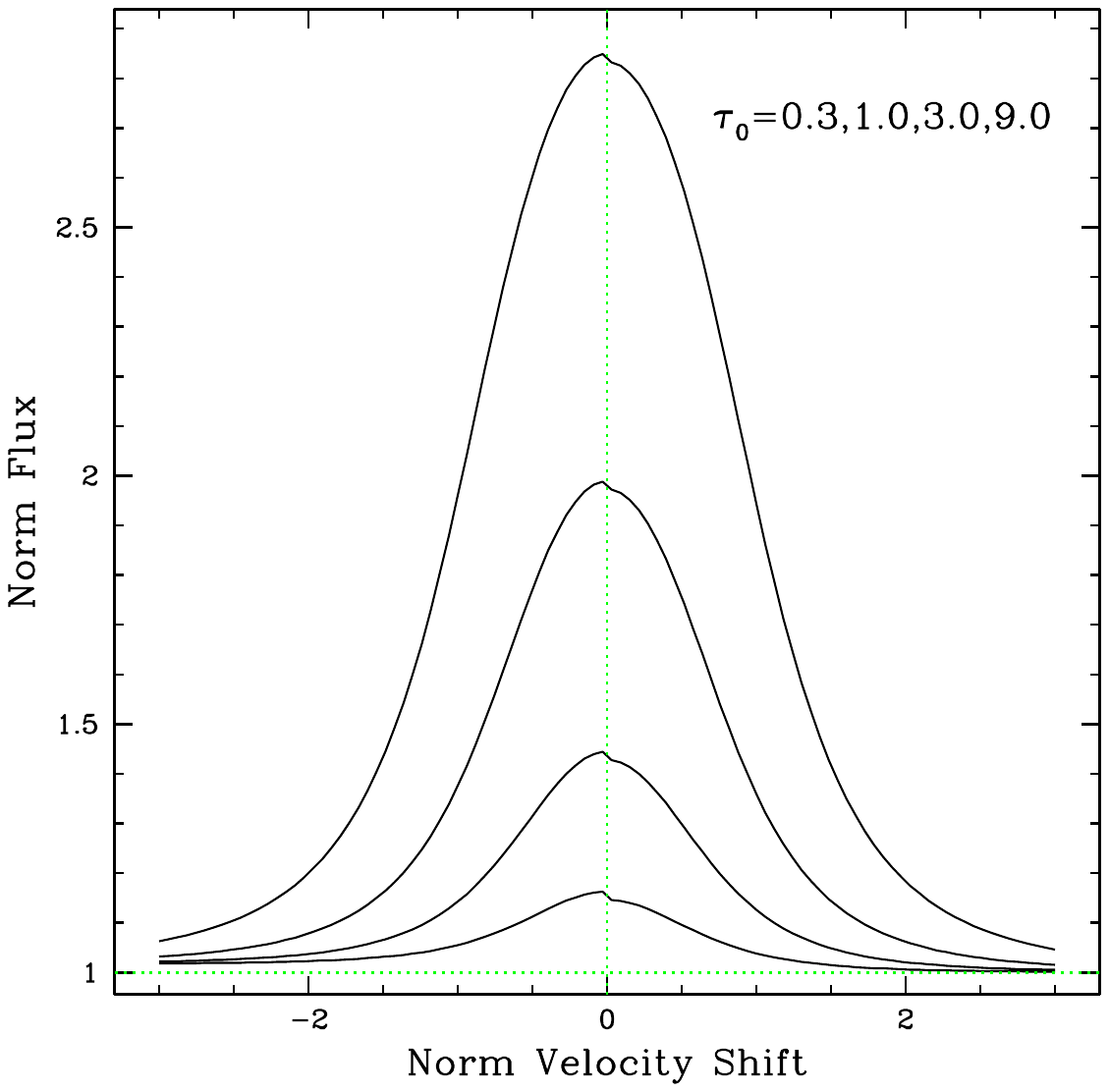}
\caption{Emission-line profiles for a spherical wind with no
CIR structure (i.e., $\eta=0$).  The wind velocity is linear with $v \propto r$, and
the density is $\rho \propto r^{-3}$.  The 4 profiles are for the
4 values of line optical depth parameter $\tau_0$ as indicated,
plotted as continuum normalized flux against normalized
velocity shift.  The vertical green line is a guide for
the location of line center.
\label{fig2}}
\end{figure}

In the spirit of proof-of-concept, we make a further simplification
by assuming $S_L=I_C$, essentially an assumption of LTE for
an isothermal wind at the temperature of the star.  While not
quantitatively accurace for WR~winds, this eliminates one further
free parameter in our study while still producing a pure
emission line as observed.  The assumption entails no net P~Cygni
absorption for blueshifted velocities. For example,
equation~(\ref{eq:intensity}) shows that $I=I_C$ when $S_L=I_C$.
This occurs because the emission from the wind exactly offsets the
absorption by the column density associated with the isovelocity
surface.

As an example, Figure~\ref{fig2} displays four recombination emission
line profiles for a purely spherical wind, so $\eta=0$ everywhere,
and we take $g(\xi)=1$.  The four line profiles are for line optical
depth parameters $\tau_0 = 0.3, 1.0, 3.0,$ and 10.0 from weakest
to strongest.  Note even though $S_L = I_C$, the emission line is
not quite symmetric because occultation by the photosphere blocks
some line emission from the hemisphere of the wind with redshifted
velocities.

\subsection{Variable Line Polarization}

Our goal is to calculate variable linear polarization across
blueshifted velocities of the line profile.  While Stokes parameter
$I$ describes the total intensity, Stokes $Q$ and $U$ describe the
linearly polarized intensity, and $V$ describes the intensity of
circular polarization.  Here we take $V=0$. To obtain a net intrinsic
polarization for an unresolved source, the source cannot be
centro-symmetric in its intensity distribution.  Our thesis is that
symmetry is broken by the presence of a CIR.  Moreover, that the
polarization shows a position angle rotation on a single spectrum
over only blueshifted velocities suggests an absorptive effect.
Specifically, axial symmetry is broken through increased
absorption of continuum limb polarization by the intervening CIR,
plus the CIR adds diluting unpolarized radiation through recombination.

The photosphere is expected to show linear polarization generally
at every point on its surface, such as limb polarization.  However,
if spherically symmetric, the net polarization will be zero. That
symmetry is broken because of differential absorption by the CIR.
The CIR stretches through the wind, threading across isovelocity
surfaces.  But those intersections can overlap, in projection
from the observer's perspective, with the photosphere generally at
different sectors with rotational phase and viewing inclination.
The result is a polarization that varies with time and across
blueshifted velocities in the line.

We adopt a centro-symmetric distribution of polarization with impact
parameter across the photosphere as an intensity of the form

\begin{equation}
I_P = G(\varpi)\,I_C(\varpi) = I_\ast\,G(\varpi)\,F(\varpi),
\end{equation}

\noindent where $G$ encodes how the amplitude of polarization varies from the center of the photosphere to the limb, and $F$ is for the corresponding limb darkening profile.  Then
the Stokes $Q$ and $U$ intensities become

\begin{eqnarray}
Q_\ast(\varpi) & = & I_P(\varpi)\,\cos 2\psi,~{\rm and} \\
U_\ast(\varpi) & = & I_P(\varpi)\,\sin 2\psi .
\end{eqnarray}

\noindent To determine limb polarization and limb darkening appropriate for a photosphere formed in an optically thick wind,
we considered spherically symmetric MCRT calculations, which are summarized in the Appendix.  For star and wind parameters similar to WR~6, $G_{\rm max} = G(\varpi=1) = 0.23$, or a maximum polarization of 23\%. Results presented in the Appendix are used in computing observables described next.

\begin{table}
\caption{Model Parameters}
\label{tab1}
\begin{tabular}{lcccccccc}
\hline
Figure & $\beta_0$ & $\eta$ & $i_0$ & $\tau_0$ & $q_C$ & $u_C$ & $q_I$ & $u_I$ \\
Figure & ($^\circ$) &  & ($^\circ$) & & (\%) & (\%) & (\%) & (\%) \\
\hline
\ref{fig3}-left  & 25 & 1 & 40 & 3 & -0.6 & -0.6 & -0.8 & +0.8 \\
\ref{fig3}-mid   & 25 & 1 & 60 & 3 & -0.6 & -0.6 & -0.8 & +0.8 \\
\ref{fig3}-right & 25 & 1 & 75 & 3 & -0.6 & -0.6 & -0.8 & +0.8 \\
\ref{fig4}-left  & 25 & 1 & 40 & 3 & -0.6 & -0.6 & -0.8 & +0.8 \\
\ref{fig4}-mid   & 25 & 1 & 60 & 3 & -0.6 & -0.6 & -0.8 & +0.8 \\
\ref{fig4}-right & 25 & 1 & 75 & 3 & -0.6 & -0.6 & -0.8 & +0.8 \\
\ref{fig5}-left  & 25 & 1 & 40 & 3 & 0.0 & 0.0 & 0.0 & 0.0 \\
\ref{fig5}-mid   & 25 & 1 & 60 & 3 & 0.0 & 0.0 & 0.0 & 0.0 \\
\ref{fig5}-right & 25 & 1 & 75 & 3 & 0.0 & 0.0 & 0.0 & 0.0 \\
\ref{fig6}-top left  & 25 & 1 & 60 & 3 & 0.0 & 0.0 & 0.0 & 0.0 \\
\ref{fig6}-bot left  & 25 & 1 & 60 & 3 & 0.0 & 0.0 & +1.0 & +2.0 \\
\ref{fig6}-top right & 25 & 1 & 60 & 3 & -0.6 & -0.6 & 0.0 & 0.0 \\
\ref{fig6}-bot right & 25 & 1 & 60 & 3 & +0.4 & -0.2 & 0.0 & 0.0 \\
\ref{fig7}-top left  & 25 & 1 & 60 & 3 & 0.0 & 0.0 & 0.0 & 0.0 \\
\ref{fig7}-bot left  & 25 & 1 & 60 & 3 & 0.0 & 0.0 & +1.0 & +2.0 \\
\ref{fig7}-top right & 25 & 1 & 60 & 3 & -0.6 & -0.6 & 0.0 & 0.0 \\
\ref{fig7}-bot right & 25 & 1 & 60 & 3 & +0.4 & -0.2 & 0.0 & 0.0 \\
\hline
\end{tabular}
\end{table}

\begin{figure}
\includegraphics[width=3.5in]{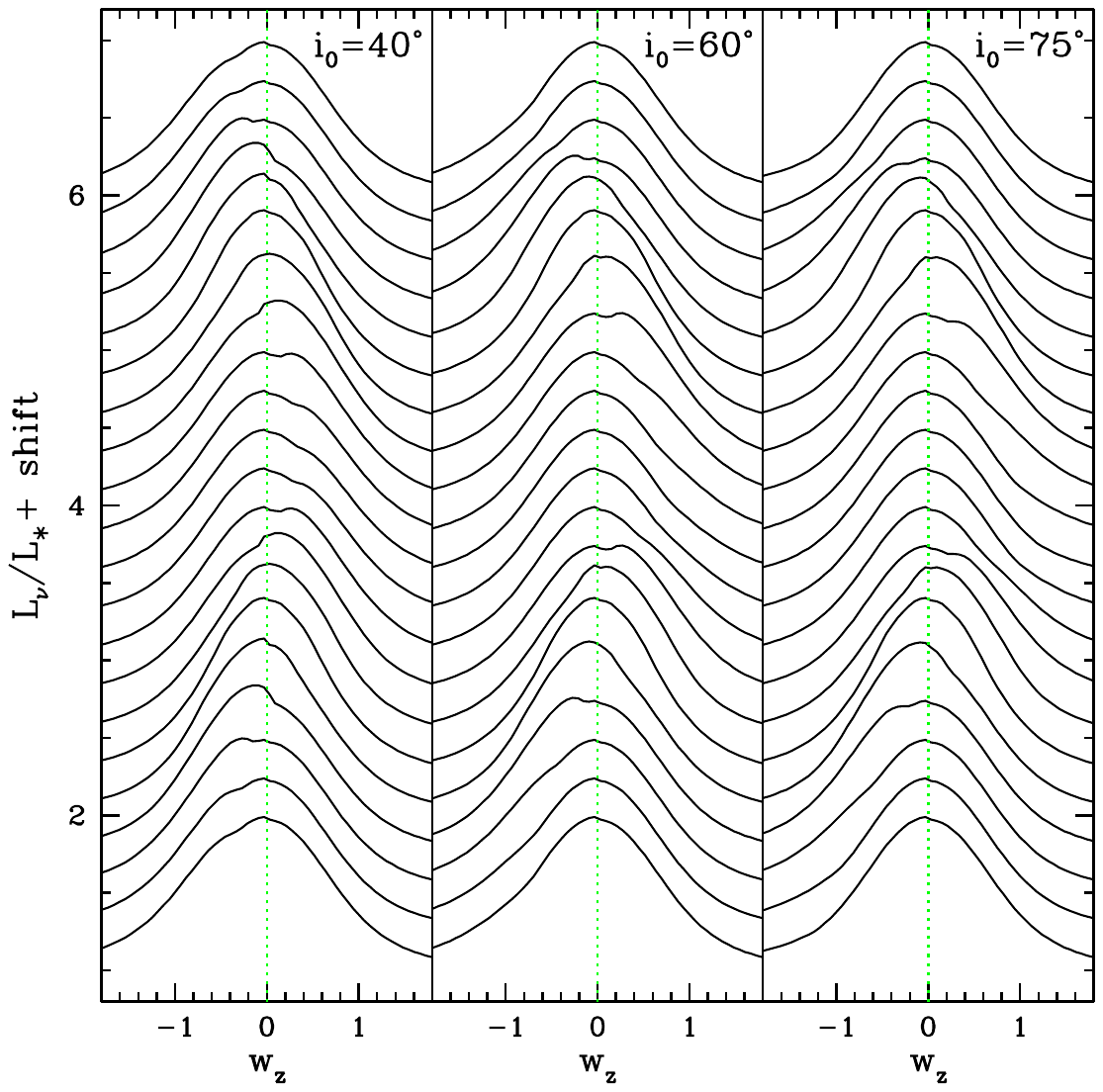}
\caption{Illustration of viewing inclination effects on the emission line profile.  The model wind has $\tau_0=3$ with a conical CIR of half opening angle $\beta_0=25^\circ$ and $\eta=1$.  The panels are for the three viewing inclinations $i_0$ as indicated.  The vertical green line is a guide for line center.  There are 21 profiles for phase intervals of 0.05 in rotation, each line being shifted slightly for viewing.
\label{fig3}}
\end{figure}

Calculation of the Stokes luminosities becomes

\begin{eqnarray}
\frac{L_Q(w_{\rm z})}{L_\ast} & = & \int\int G(\varpi)F(\varpi)\cos 2\psi\,
	e^{-\tau_L}\,\varpi d\varpi\frac{d\psi}{\pi},~{\rm and} \\
\frac{L_U(w_{\rm z})}{L_\ast} & = & \int\int G(\varpi)F(\varpi)\sin 2\psi\,
	e^{-\tau_L}\,\varpi d\varpi\frac{d\psi}{\pi}.
\end{eqnarray}

\noindent The net polarization arises entirely from attentuation
by destructive absorption of the continuum polarization.  Note that
additional polarization from scattered light in the asymmetric wind
is expected \citep[c.f.,][]{2015A&A...575A.129I}, but its influence
would not be limited to blueshifted velocities only.  We subsume
such an effect in the free parameters for the continuum polarization (see
next section) so we may focus on polarimetric behavior limited to
only blueshifted velocities in the line.

\section{Parameter Study}
\label{sec:study}

For our parameter study, we adopt the following assumptions, several of
which have been noted already:  

\begin{enumerate}

\item The CIR is conical with half-opening angle $\beta_0$.

\item The CIR stems from the equator.

\item When a point lies within the CIR, the density is enhanced by
the factor $1+\eta$ with $\eta > 1$. 

\item The velocity field inside the CIR is identical to that
of the otherwise spherical wind.

\end{enumerate}

\noindent Additionally, as a matter of convention, the angle $\psi$
is defined such that $Q>0$ for polarization parallel to the rotation
axis and $Q<0$ for polarization perpendicular to that axis.

The free parameters of the model consist of the following

\begin{enumerate}

\item The viewing inclination $i_0$ which is zero for a pole-on
view of the stellar rotation axis and $90^\circ$ for an
equator-on view.

\item The line optical depth scale $\tau_L$.

\item The half opening angle $\beta_0$ of the CIR cone.

\item The density contrast of the CIR relative to the otherwise spherical wind, $\eta$.

\end{enumerate}

\begin{figure}
\includegraphics[width=3.5in]{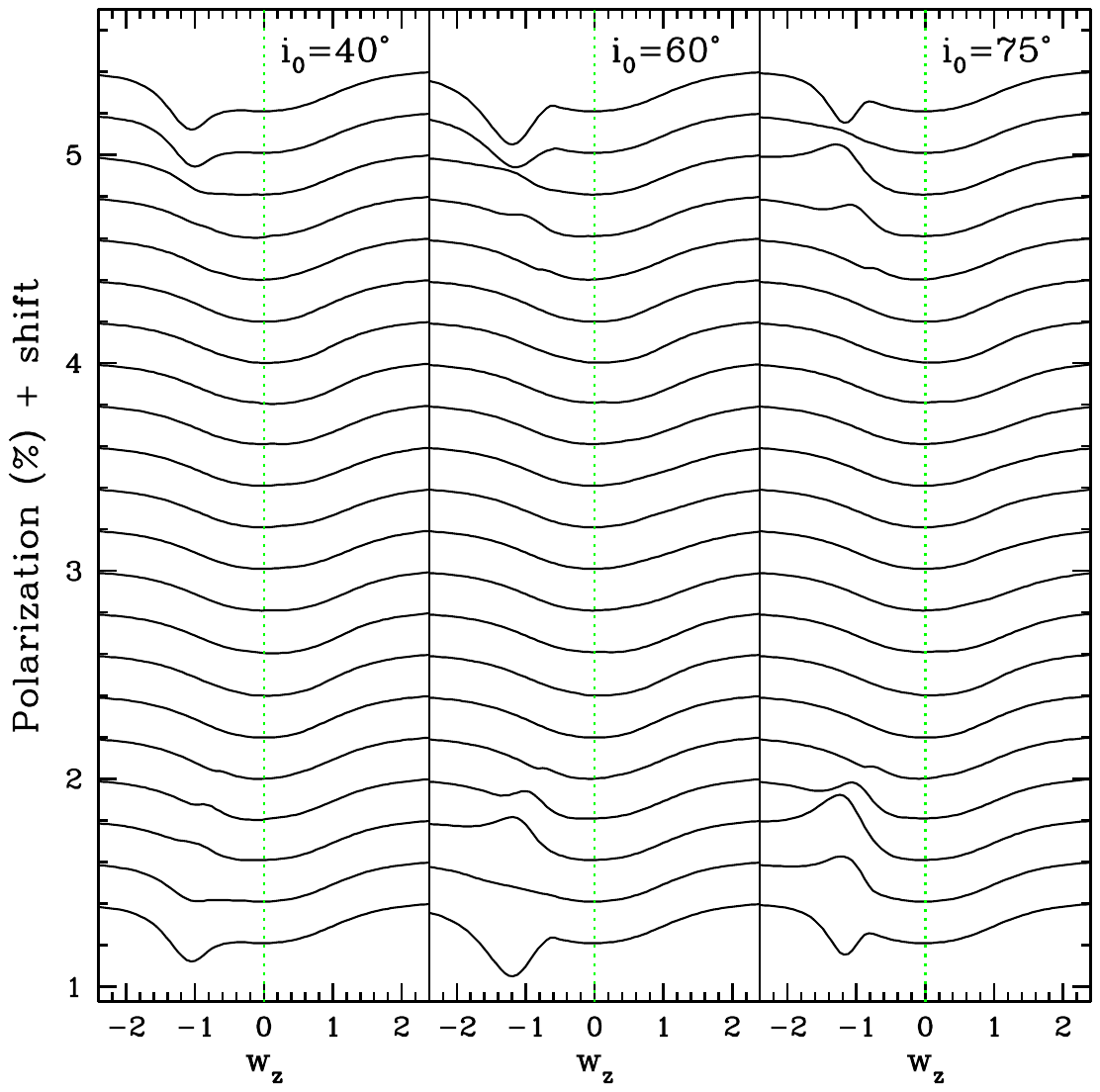}
\caption{Similar to Fig.~\ref{fig3}, now for the polarization across the emission line.  Here the continuum polarization is taken to have Stokes parameters of $q_C=-0.6\%$ and $u_C=-0.6\%$, similar to the blue continuum point in Fig.~\ref{fig1}.  There is a depressed level of polarization across the central part of the polarized line profile owing to the line effect.  There are also additional features in the blue wing when the CIR is in front of the star.  These effects are present for the first quarter and last quarter of the curves, but not the middle half of them when the CIR rotates behind the star.
\label{fig4}}
\end{figure}

\begin{figure}
\includegraphics[width=3.5in]{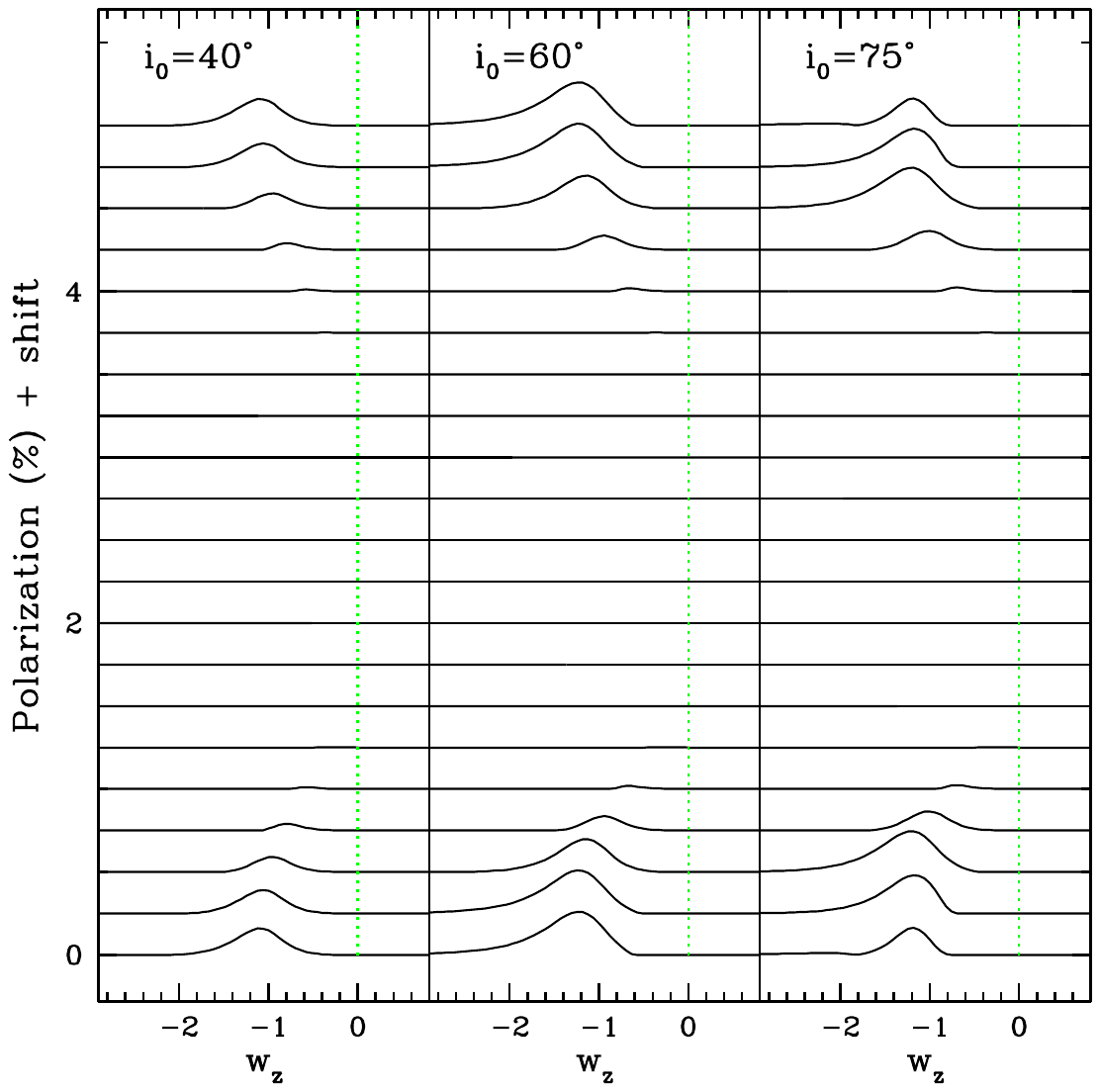}
\caption{The same models as Figs.~\ref{fig3} and \ref{fig4}, now
assuming no interstellar or intrinsic continuum polarizations (i.e.,
$q_I=u_I=q_C=u_C=0$).  The removal of continuum polarization eliminates
the polarimetric dilution of the emission line while isolating the
effect by the CIR.  The vertical green line is a visual guide for
line center (note that velocity shift is not shown symmetric about
the line since the polarization is zero at redshifted velocities).
\label{fig5}}
\end{figure}

As an additional simplication, in equation~(\ref{eq:kappa}), we assume $g=1$ for all the model results shown here,
in which case $\tau_L \propto \rho^2$.  The star rotates with period
$T_\ast$ and angular speed $\omega = 2\pi/T_\ast$, such that the
azimuth for the center of the CIR at the equator is $\phi_0(t) =
\omega\, t$.  Finally, our models only allow for one CIR to highlight the effect of interest.

The deficiencies of the model are clear.  The CIR need not be
conical but could be spiral \citep[e.g.,][]{1996ApJ...462..469C}.  The CIR need not be at the equator \citep[c.f.,][]{2015A&A...575A.129I}.
There could be multiple CIRs \citep[e.g., the BRITE study of $\zeta$ Pup with two CIRs,][]{2018MNRAS.473.5532R}.  The density distribution in the CIR
need not mimic that of the spherical wind, plus the velocity field
in and around the CIR can be non-monotonic \citep{2017MNRAS.470.3672D}.  Our objective here is proof-of-concept not fits to observations, which would require a far more tailored analysis.

With these caveats in mind, we do wish to include some practical
issues pertaining to measurements.  We allow for interstellar
polarization and intrinsic continuum polarization in the following
way.  The interstellar polarization is chromatic, but its position
angle is constant.  Its wavelength dependence is slow and can be
taken as constant across any particular wind-broadened line.  We assign variables
$q_I$ and $u_I$ for the relative polarizations of the Stokes
parameters.  For example, for a spherical wind with no intrinsic
polarization, polarized fluxes of $f_Q = q_I \, f_\ast$ and $f_U =
u_I\, f_\ast$ would be measured at Earth with polarization position
angle $\tan(2\psi_I) = u_I/q_I$.

For the intrinsic continuum polarization, we refer to values of
$q_C$ and $u_C$ that would be measured outside the line.  We
acknowledge that with a CIR, some degree of polarization in the
continuum is expected, and that polarimetric variability would be cyclic with rotational
phase.  We do not model the continuum polarization self-consistently but allow for it as
we anticipate a ``line effect'' that depresses the continuum
polarization within the line \citep{1990ApJ...365L..19S}.  This is a well-known effect and
occurs when emission line flux is not scattered but acts only to
dilute the polarization.  It is an artifact of the definition of
the polarization as a ratio of polarized flux to total flux.  In
the continuum the relative polarization is $q_C$ and $u_C$.  But
within the line, if line photons are not scattered, the relative
polarizations are depressed by the factor $L_C/(L_C+L_L)$, where
$L_C$ is the continuum luminosity outside the line and within the line $L_L=L_I-L_C$, for $L_I$ from equation~(\ref{eq:LI}).

The final polarizations become

\begin{eqnarray}
q(t) & = & q_L(t,w_{\rm z}) + q_I + q_C(t)\cdot\left[\frac{L_C}{L_C+L_L(t,w_{\rm z})}\right]~{\rm and}, \\
u(t) & = & u_L(t,w_{\rm z}) + u_I + u_C(t)\cdot\left[\frac{L_C}{L_C+L_L(t,w_{\rm z})}\right],
\end{eqnarray}

\noindent where $q_L$ and $u_L$ represent the relative Stokes
polarizations as given by

\begin{eqnarray}
q_L(t) & = & \frac{L_Q(t,w_{\rm z})}{L_C+L_L(t,w_{\rm z})},~{\rm and} \\
u_L(t) & = & \frac{L_U(t,w_{\rm z})}{L_C+L_L(t,w_{\rm z})}.
\end{eqnarray}

\begin{figure}
\includegraphics[width=3.5in]{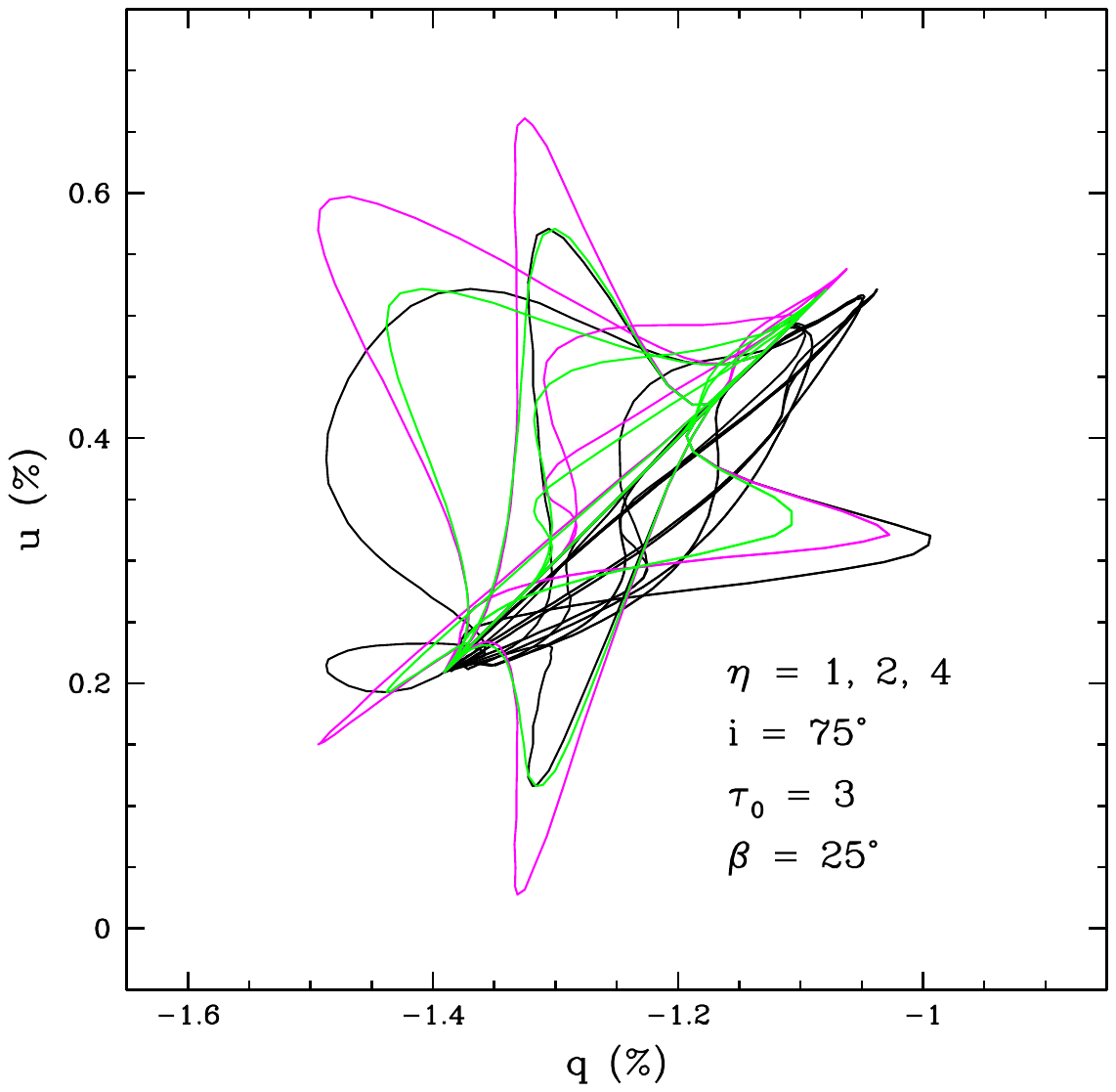}
\caption{A $q-u$ figure with model parameters as shown and emphasizing how the value of $\eta$ alters the pattern of polarimetric variability.  The colors are $\eta=1$ (green), 2 (magenta), and 4 (black).  Note that $q_C=u_C=-0.6\%$, $q_I=-0.8\%$, and $U_I=+0.8\%$.  Each color represents 21 rotational phases, from 0.0 to 1.0 (with 0.0 and 1.0 being degenerate); however, only about half those phases produce loops corresponding to when the CIR is in front of the star.
\label{fig8}}
\end{figure}

\begin{figure}
\includegraphics[width=3.5in]{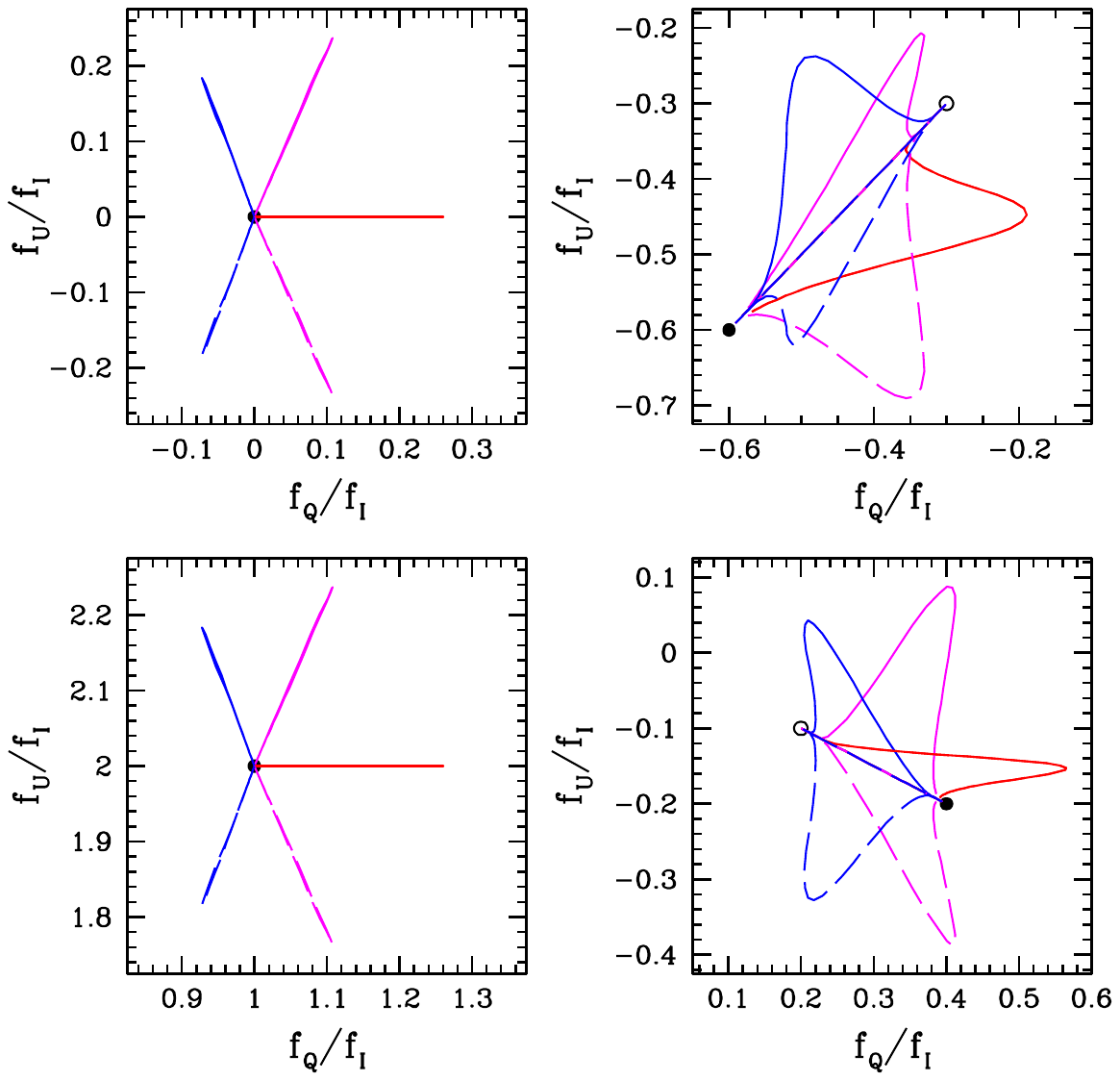}
\caption{Illustrative cases for $q-u$ variations for different sets of model parameters (see Tab.~\ref{tab1}).  A given curve is the polarization across the line profile from blue wing through line center across the red wing.  Left panels have no intrinsic polarization in the source continuum and show a spoke-like pattern.  Upper left has no interstellar polarization, whereas lower left does.  Right panels do have continuum polarization but no interstellar polarization.  The linear portions are for the red wing (changing polarization owing to the line effect, but no position angle rotation), whereas the loop discursions are all from the blue wing.
\label{fig6}}
\end{figure}

\begin{figure}
\includegraphics[width=3.5in]{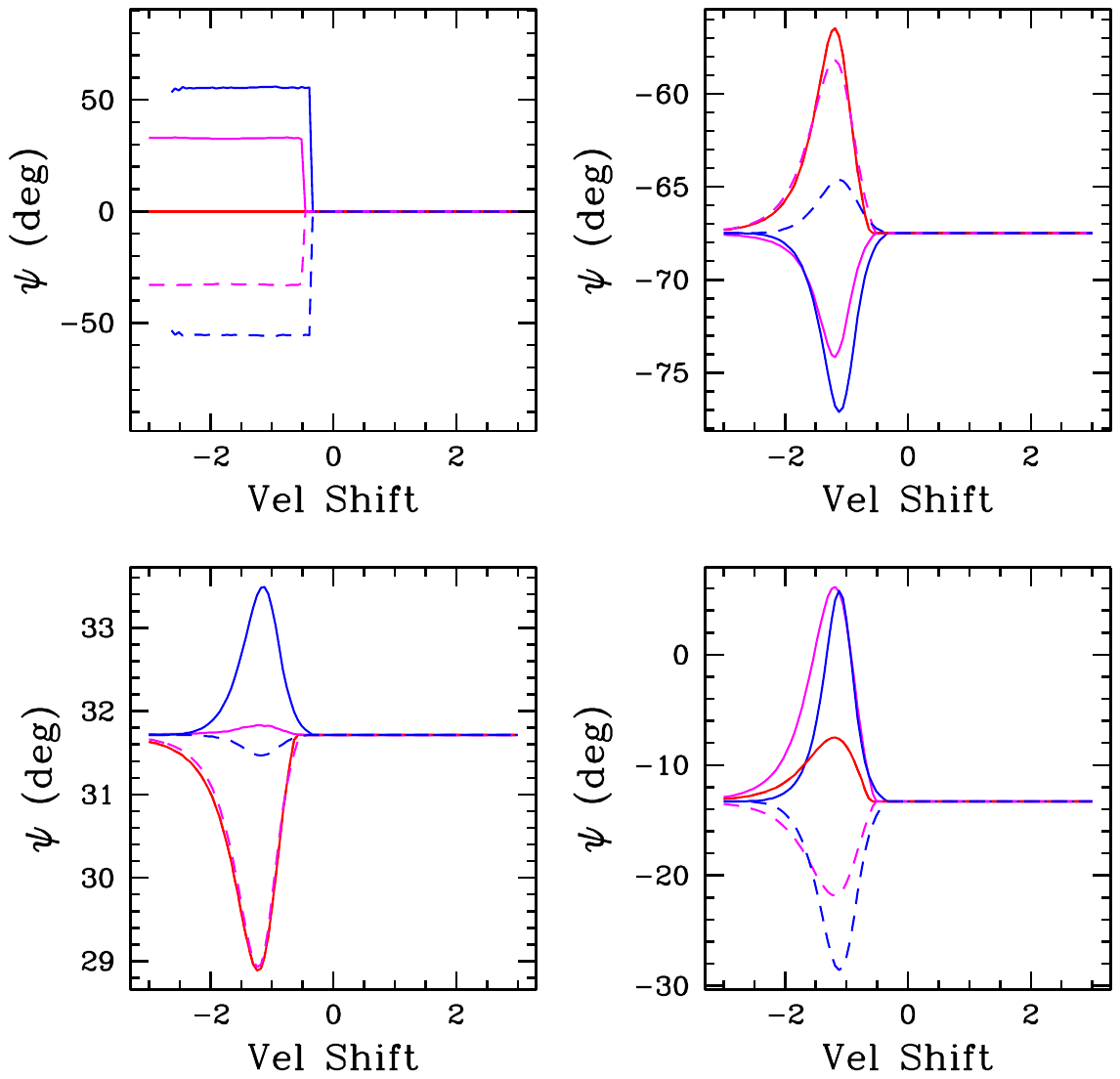}
\caption{Same models as shown in Fig.~\ref{fig6}, now plotting specifically the polarization position angle $\psi$ against normalized velocity shift.  With no continuum polarization, upper left shows only horizontal lines.  Different curves are for different rotational phases.  The position angle signifies the portion of the sky masked by the overdense conical CIR structure against the star.  The one horizontal line extending to redshifts is the unvarying polarization position angle reflective of the continuum and/or interstellar medium parameters.
\label{fig7}}
\end{figure}

\noindent Note that $q_L$ and $u_L$ arise from absorption, so affect
only the blueshifted velocities.  By contrast the line effect
influences both red and blue shifts, and it goes away outside the
line.  Also, factors $q_C$ and $u_C$ can be functions of time, such as when the continuum polarization arises from scattered light by the CIR \citep[c.f.,][]{2015A&A...575A.129I, 2018MNRAS.474.1886S}.
Model results are illustrated in Figures \ref{fig3}--\ref{fig8}.  For the multi-panel figures \ref{fig3}--\ref{fig7}, model parameter selections are detailed in Table~\ref{tab1}.

In Figures \ref{fig3} and \ref{fig4}, the effects of the CIR on the line profile shape and polarization are shown for 3 different viewing inclination angles.  Emission profile variations are shown in Figure \ref{fig3} for 21 equally spaced rotational phases, from bottom (phase 0) to top (phase 1, which is the same as phase~0).  The profiles are continuum normalized and plotted against normalized velocity shift $w_{\rm z}$. Each profile is vertically shifted by a small amount for ease of viewing.  The selection of viewing inclinations divides the solid angle of the sky, from the point of view of the star, into 4 equal solid portions.  Consequently for a random observer, the odds of viewing the CIR poleward of $40^\circ$, between $40^\circ-60^\circ$, between $30^\circ-75^\circ$, and equatorward of $75^\circ$ are all equal.  (Likewise for the opposite hemisphere.)

The polarization across the profiles are shown in the same fashion in 
Figure~\ref{fig4}.  Continuum polarization values for $q_C$ and $u_C$ are chosen that are similar to the blue continuum point observed for WR~6 as shown in Figure~\ref{fig1}.  Additionally, non-zero values of interstellar polarization values were assigned as $q_I$ and $u_I$ as well.  The presence of continuum polarization intrinsic to the star results in the classic ``line effect'', whereby polarization is depressed inversely to the line emission.  In addition, the conical CIR imposes variable polarization only at blueshifts ($w_{\rm z}<0$) and when forefront of the star, corresponding to the lower five and upper five line profiles.

The same CIR model is shown in Figure~\ref{fig5}, now with no polarization arising from either the source continuum nor imposed by the interstellar medium.  Here the effects of the CIR are dramatically evident.  Note that the panels emphasize blueshifts, showing little of the red wings which are just flat lines of zero polarization.  The vertical green line is a  for line center.  

In this figure, as expected, there is no line effect when there is no continuum polarization.  The bumps in the polarization arise entirely from the CIR producing enhanced absorption of the limb polarization of the stellar pseudo-photosphere.  The location of the bump drifts in velocity shift as the CIR structure rotates about the star.  The amplitude changes as well.  These effects arise because of how the conical CIR intersects regions of isovelocity surfaces that lie in front of the star.  Changing the inclination alters the strength of the bump.  When the CIR rotates to the back side of the star, no bumps are present.

To illustrate the effect of density contrast in the CIR as compared to the otherwise spherical wind, Figure~\ref{fig8} shows 4 models in the $q-u$ plane.  Model parameters are indicated in the plot space, with $\eta$ values of 1 (green), 2 (magenta), 3 (blue), and 4 (black).  Each complete loop is for a different phase, with of course only about half the phases producing loops (i.e., when the CIR is forefront of the star).  The patterns are all qualitatively similar, mainly with reduced amplitude of variation with lower $\eta$.

Figure~\ref{fig8} is a ``mashup'' of many phases.  The next two figures highlight a few phases for greater clarity, and also indicate that the model can produce both clockwise (CW) and counterclockwise (CCW) senses of PA rotation.  The sense of PA rotation is with reference to velocity shifts, meaning as the polarization and PA varies mostly smoothly from line center to continuum, what is the handedness of that variation.

Figures~\ref{fig6} and \ref{fig7} display a 4-panel set of models for different combinations of continuum and interstellar polarization values and only select rotational phases of 0.00 (red), 0.05 (magenta), 0.10 (blue) in solid and 0.90 (blue), 0.95 (magenta), and 1.00 (red) in dashed.  A reminder that at phase 0.00 and 1.00, the CIR is directly forefront of the star (but at some viewing inclination).  Phase 0.25 is when the CIR has rotated to the plane of sky, right of the star in projection by our convention; then phase 0.75 is when the CIR has rotated fully behind the star to again be in the plane of the sky, now left side of the star.

The same models are displayed in Figures~\ref{fig6} and \ref{fig7}, where the former shows line profiles displayed in the $q-u$ plane, and the latter shows polarization position angle, $\psi$, variations with velocity shift in the lines.  Model parameters are provided in Table~\ref{tab1}.  For Figure~\ref{fig6}, the two left panels have zero continuum polarization; upper has no interstellar while lower does have interstellar.  The two panels at right both have continuum polarizations (different values of $q_C$ and $u_C$), but neither has interstellar polarization.  When there is no continuum polarization, the $q-u$ patterns are ``spokes'' (left); where there is continuum polarization, loop patterns are apparent for the blue wing, but linear variations appear for the red wing.

For Figure~\ref{fig7}, the patterns seen in Figure~\ref{fig6} map into $\psi$ versus $w_{\rm z}$ as fixed PAs on the red wing, but generally variable PAs on the blue wing.  One exception is the upper left panel. In the absence of continuum polarization, each rotation phase is a constant PA on the blue wing, but that PA differs from the always fixed PA for the red wing.  Note the small rotation in PA for the lower left panel.  Here there is no continuum polarization, but there is interstellar polarization.  Since the line polarization itself varies with velocity shift, and since the PA is defined as $\tan 2\psi = u_{\rm tot}/q_{\rm tot}$, a small PA results.  Whether the effect is large or small depends on how the interstellar polarization compares with that from the line.

%\begin{figure}
%\includegraphics[width=3.5in]{syst01.pdf}
%\caption{\label{fig7}}
%\end{figure}
%\begin{figure}
%\includegraphics[width=3.5in]{syst02.pdf}
%\caption{\label{fig8}}
%\end{figure}
%\begin{figure}
%\includegraphics[width=3.5in]{syst03.pdf}
%\caption{\label{fig9}}
%\end{figure}
%\begin{figure}
%\includegraphics[width=3.5in]{syst04.pdf}
%\caption{\label{fig10}}
%\end{figure}

\section{Data Availability}

No new data were generated or analysed in support of this research.

\section{Summary}
\label{sec:summary}

The most massive stars experience substantial mass-loss rates over their stellar lifetimes, while both on the main sequence and during post-main sequence stages (such as LBV and WR phases).  Accurate mass-loss rates are needed to constrain stellar atmosphere and evolution models that predict ionizing radiation, properties of progenitors of supernovae, rotational histories, and the properties of remnants.  However, structure in the wind outflows create uncertainties in many observational wind diagnostics.  

This structure takes two primary forms:  stochastic such as clumping and organized such as CIRs.  The latter are coherent features that thread the wind and produce periodic or quasi-periodic effects tied to the rotation of the star.  While CIRs are inferred from the absorption troughs of P Cygni UV resonance lines for OB stars, their presence has been less studied for WR winds at UV wavelengths \citep[see][for one example]{1992A&A...266..377P}.  However, evidence for their presence has been building for the WR class.  Since the hydrostatic layers are unobservable owing to the optically thick winds, CIRs are critically important for providing rotation periods of WR stars, for which there seems few other means for measuring the periods.

While CIRs can produce a variety of observable effects, the one
emphasized in this paper is the prospect of variable polarization
across emission lines that occurs at blueshifted velocities but
not redshifted ones.  We developed a simplistic model to explore
this diagnostic consisting of a conical CIR in an otherwise spherical
wind, and a radially expanding flow following a linear velocity law
with $v \propto r$.

While such a model does not yield quantitatively accurate results, our simulations provide qualitative insights relevant to observations.  (1) For a pseudo-photosphere formed in the wind with a distribution of limb polarization, a CIR can indeed produce time-variable polarization restricted to blueshifted velocities when the CIR is forefront of the star.  (2) The CIR produces no such effects when it rotates to the backside of the star.  (3) Different loop patterns in the $q-u$ plane result as a function of the level of continuum polarization outside the emission line.  

These general features should be fairly robust for more realistic wind and CIR models.  It is certainly possible that more realistic velocity fields inside and outside the CIR, combined with a rotational component and with possible spiral morphology of the CIR will lead to a richer spectrum of variable behavior.  In particular, some variable polarization could appear at redshifted velocities, but likely closer to line center.  

Archival data of linear spectropolarimetry of the He{\sc ii} 4686 line from WR~6 obtained with HPOL in the 1990s hint at the effects explored here.  Data of higher precision and spectral resolution from ESPaDOnS display several of the effects have been modeled and will appear in a future paper (Fabiani et~al, in prep).

\section*{Acknowledgements}

The authors express appreciation to Ken Gayley for insightful
and encouraging remarks that have improved this manuscript.  RI and
CE gratefully acknowledge that this material is based upon work
supported by the National Science Foundation under grant number
AST-2009412. The work of ANC is supported by NOIRLab, which is
managed by the Association of Universities for Research in Astronomy
(AURA) under a cooperative agreement with the National Science
Foundation.  NSL and AFJM acknowledge financial support from the
National Sciences and Engineering Council (NSERC) of Canada.

\bibliographystyle{mnras}
\bibliography{ignace} % if your bibtex file is called example.bib

\appendix

\section{Limb Darkening and Limb Polarization for a Spherical
and Optically Thick Wind}
\label{sec:app}

We employed the Monte Carlo radiative transfer (MCRT) routines of \cite{1996ApJ...461..828W}
to simulate the scattered light distribution of a spherically
symmetric and electron-scattering wind.  With linear expansion 
$v(r) \sim r$, the wind density is $\rho \sim r^{-3}$.  The 
line-of-sight optical depth to the wind base from the observer 
in the direction of the center of the star is

\begin{equation}
    \tau = \int_{R_\ast}^\infty\, \sigma_T\,n_{\rm e}(r)\,dr
    = \tau_0\,\int_0^1\,\xi\,d\xi = \tau_0/2,
\end{equation}

\noindent where we have used $\xi=R_\ast/r$.  

\begin{figure}
\includegraphics[width=3.5in]{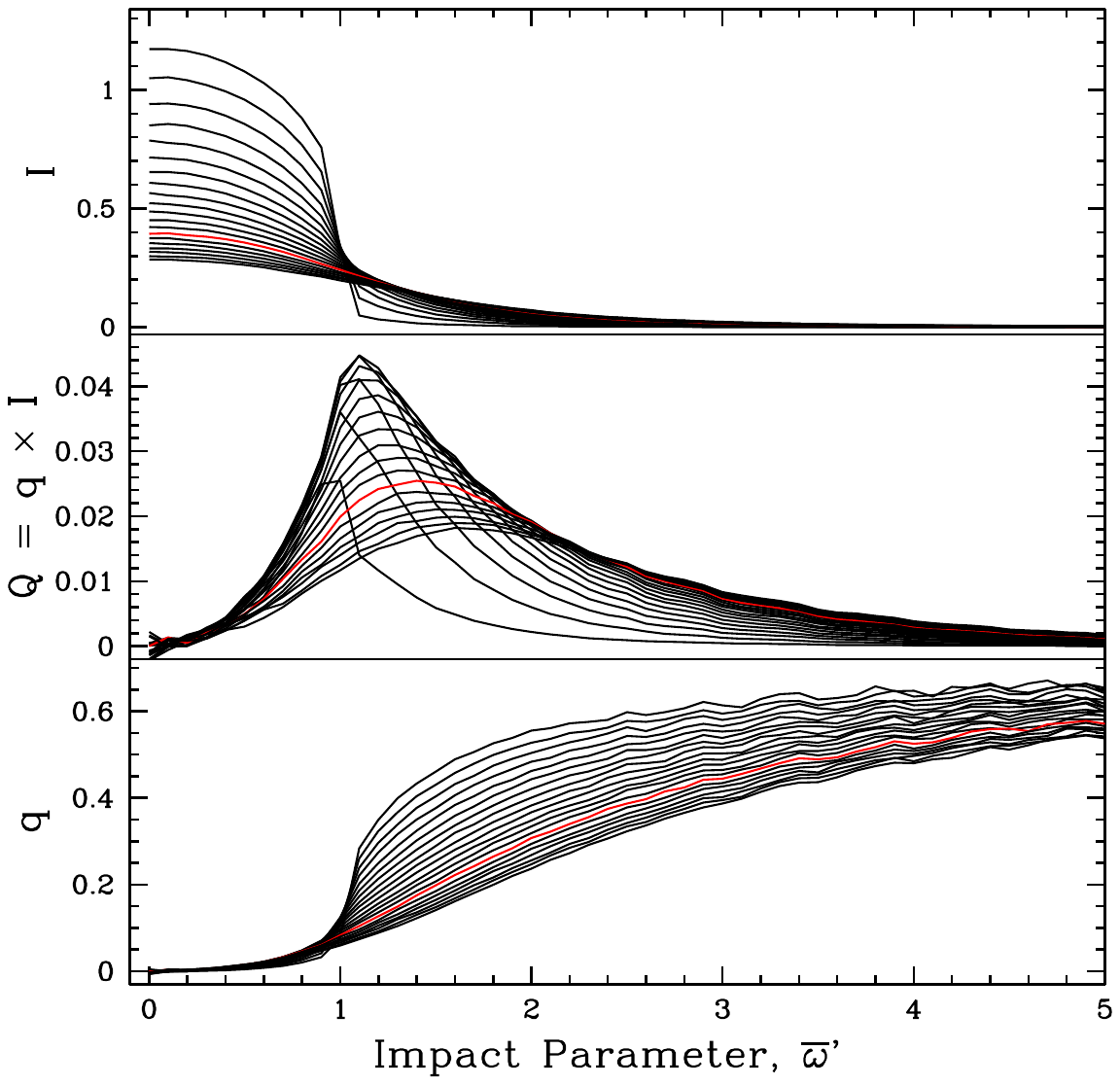}
\caption{The results of MCRT radiative transfer models for a spherically symmetric wind with inverse cube density and pure electron scattering.  Panels are emergent Stokes-I (top), Stokes-Q (middle), and relative polarization $q$ (bottom), plotted against observer impact parameter $\varpi'$ relative to the stellar radius (i.e., $\varpi'=1$ corresponds to $R_\ast$).  Each curve is for a different optical depth from 0.1 to 3.9 in steps of 0.2.  The red curve is for $\tau=2.7$ corresponding to WR~6.  These models allow for limb polarization and limb darkening of the hydrostatic atmosphere.
\label{figA1}}
\end{figure}

We ran a suite of
models for total optical depths of $\tau=0.1$ to 3.9 in steps of 0.2
as shown in Figure~\ref{figA1}.  Note that $\tau_0=2\tau$.
These panels show the emergent total Stokes $I$ intensity, Stokes $Q$ intensity, and relative polarization as $q=Q/I$ with lines-of-sight at impact parameter $\varpi'$.  Note that $\varpi'$ is normalized to $R_\ast$ (hence $\varpi'=1$ is the stellar limb).  By contrast, $\varpi$ (without a prime) of earlier sections was normalized to $R_{\rm phot}$.  While the net flux of polarization cancels identically owing to symmetry, the intensity emerging from any ray through the wind is generally linearly polarized.  It is the asymmetric absorption of this background polarization on the photosphere by the CIR that leads to effects described in this paper.

\begin{figure}
\includegraphics[width=3.5in]{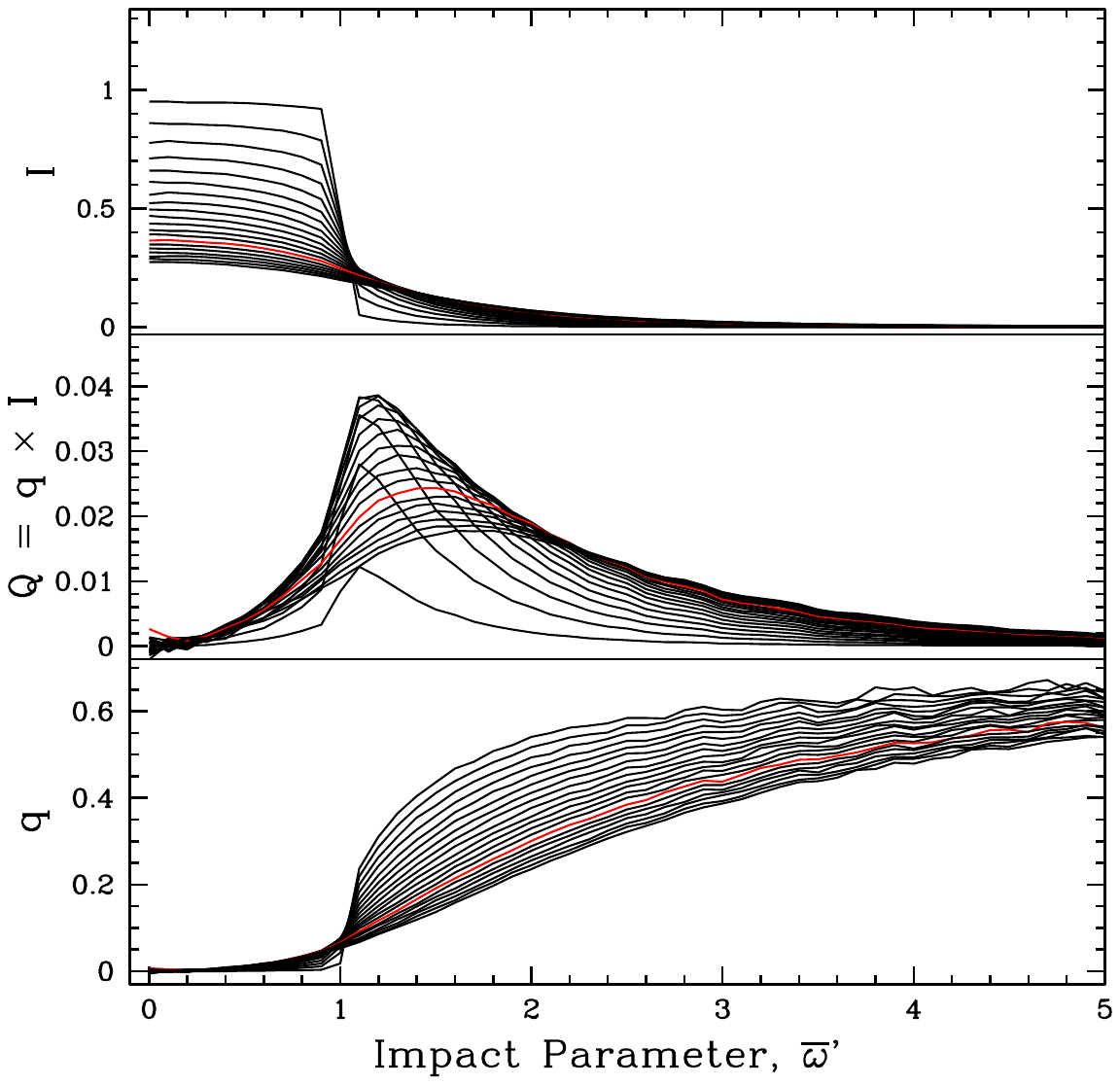}
\caption{The same as Fig.~\ref{figA1} but for a star with no limb polarization and no limb darkening (i.e., the intensity is taken as uniform at the wind base).
\label{figA2}}
\end{figure}

\begin{figure}
\includegraphics[width=3.5in]{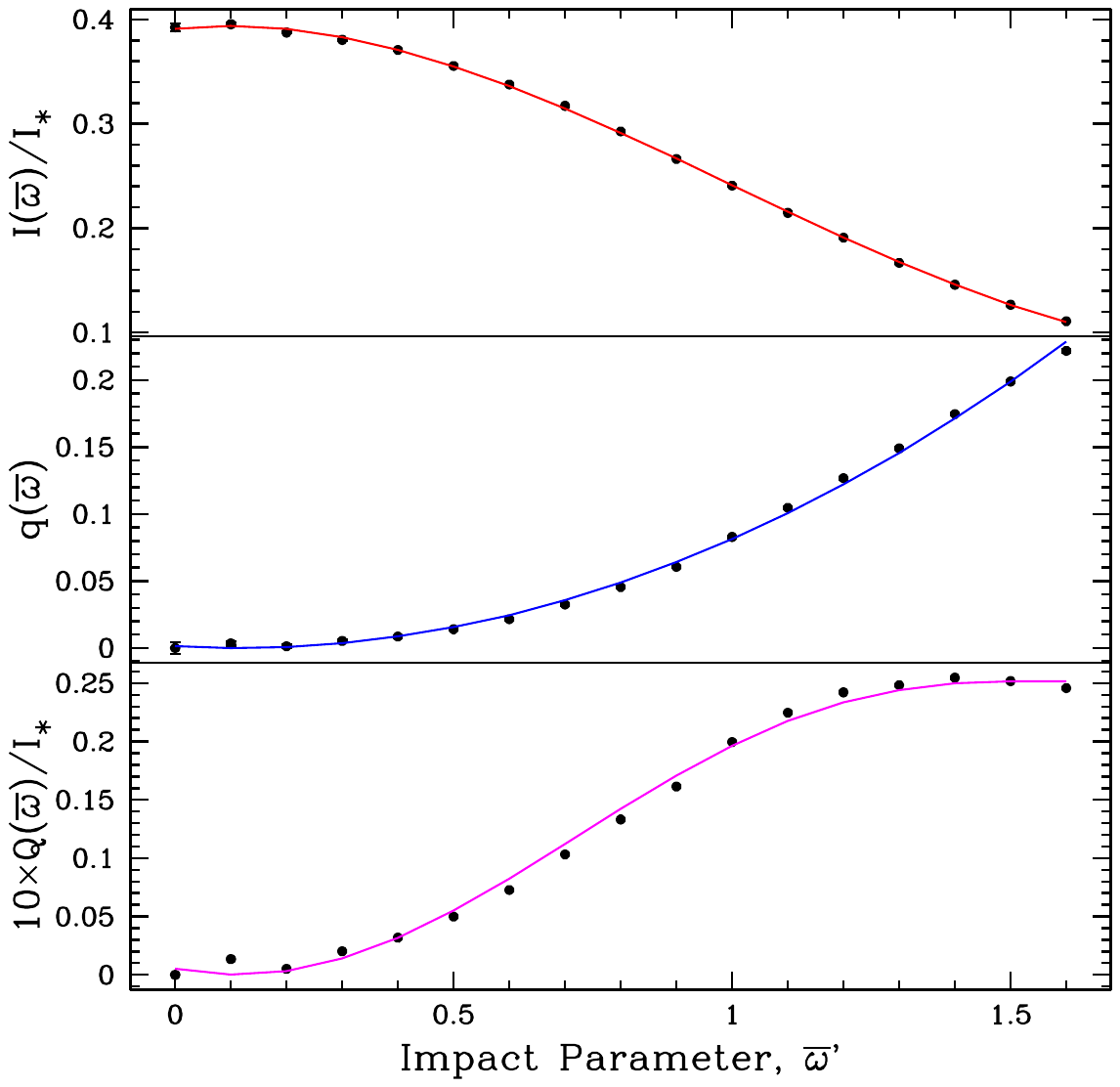}
\caption{Using the red curve for $\tau=2.7$ from Fig.~\ref{figA1}, fits are made to the MCRT results (shown as points) up to the impact parameter where Stokes-Q peaks, at $\varpi' \approx 1.6$, for Stokes-I (top; red) and $q$ (middle; blue).  Then the fit for $Q$ is implied as product of the fits with $q \times I$ (bottom; magenta). The fit formulae 
are given in Eqs~(\ref{eq:figI})--(\ref{eq:figq}).
\label{figA3}}
\end{figure}

Note that the $I$ and $Q$ intensities are relative to the case of no wind (i.e., $\tau=0$), no limb darkening, nor limb polarization.  A bare star that is uniformly bright is taken to have stellar intensity $I_\ast = 1$.  In Figure~\ref{figA1}, the hydrostatic star has a standard limb polarization and limb darkening profile of a plane-parallel atmosphere solution for pure scattering 
\citep[e.g.,][]{1960ratr.book.....C}.
By contrast Figure~\ref{figA2} shows the set of runs now for a star with neither limb polarization nor limb darkening.  This is made clear by the fact that at low wind optical depth, the intensity profile from $\varpi=0$ to 1 is flat with $I=I_\ast=1$.

As the optical depth is increased, starlight is redistributed over a larger range of impact parameters as conservative and gray electron scattering leads to multiple scattering and diffusion of light over an increasingly large volume about the star.  The result is an ill-defined effective pseudo-photosphere.  Here ``ill-defined'' refers to the idea of a breakdown in the standard core-halo approach, where the photosphere can be identified with a hydrostatic atmosphere, or``core'', and the circumstellar environment is the ``halo''. In impact parameter there is no longer a clean transition from rays that intercept a thick atmosphere to those that pass through a thin wind.  However, it is clear that once the wind becomes optically thick to scattering, the emergent $I$ and $Q$ distributions no longer depend on the details of the intensity distribution at the hydrostatic layers of the wind base, namely whether or not it has limb darkening or limb polarization.  

Based on \cite{2018MNRAS.474.1886S}, the optical depth for the wind of WR~6 is about $\tau=2.7$, which appears as the red curve in Figures~\ref{figA1} and \ref{figA2}.  Figure~\ref{figA3} shows this model singled out over a more restricted range of impact parameters.  Figure~\ref{figA3} also shows a fit to the intensity $I$ and to the relative polarization $q$.  The fit applies to $\varpi' \le 1.6R_\ast$ corresponding to the peak in the Stokes intensity $Q$.  In our proof-of-concept approach, we retain the core-halo approach for conenience, and we use the following fit formulae to model the limb polarization and limb darkening of the effective pseudo-photosphere:

\begin{eqnarray}
I/I_\ast & = & 0.39+0.06\varpi' -0.31\varpi'^2+0.10\varpi'^3 \equiv F(\varpi), \label{eq:figI} \\
q & = & -0.023\varpi' +0.103\varpi'^2 \equiv G(\varpi), \label{eq:figq} \\
Q/I_\ast & = & q \times I = G(\varpi)\cdot F(\varpi),. \label{eq:figQ}
\end{eqnarray}

\noindent where in the last equation, the implied fit is a quintic order polynomial.  Again, $\varpi = 1.6\varpi'$, since $\varpi'$ is relative to $R_\ast$, whereas $R_{\rm phot}=1.6R_\ast$ was adopted as the radial extent of the pseudo-photosphere for WR~6 as our example. 

The preceding expressions are used for the modeling of the variable emission line profile variability in $I$, $Q$, and $U$ fluxes described in this study treating the pseudo-photosphere as if it where a spherical atmosphere.  In terms of the model output, the peak intensity for the case of $\tau=2.7$ is $I(\varpi'=0) = 0.39 I_\ast$.  The intensity at the limb of the pseudo-photosphere is $I(\varpi'=1.6)=0.11I_\ast$.  The limb darkening is well fit by a cubic polynomial in $\varpi$ with a limb that is $0.11/0.39=28\%$ as bright as the central intensity.  The standard limb darkening result of an Eddington gray plane-parallel atmosphere is $2/5 = 40\%$ for the emergent intensity parallel to the atmosphere relative to normal to the atmosphere.  As expected, sphericity effects lead to stronger limb darkening \citep{2011A&A...530A..65N, 2012IAUS..282..243N}.  Using the fit expression, the specific stellar luminosity associated with the pseudo-photosphere is

\begin{equation}
L_\ast = 8\pi^2\,R_\ast^2\times 0.58\,I_\ast 
    = 8\pi^2\,R_{\rm phot}^2\times 0.23\,I_\ast
\end{equation}

\noindent where the factor of 0.58 derives from the fit formula, and the factor of 0.23 derives from recasting the expression in terms of $R_{\rm phot} = 1.6 R_\ast$.

Note also that the limb polarization is $q(\varpi'=1.6) = 0.23$ or 23\%, much higher than the classic 11\% of the plane parallel result by \cite{1960mes..book.....S}.  Again, higher limb polarization is the result of sphericity effects combined with a power-law wind density \citep{1971MNRAS.154....9C}.  The variation of $q$ with $\varpi$ is well fit by a quadratic polynomial.  These fits were adopted for the pseudo-photosphere brightness variations in Stokes $I$ and $Q$ for the line profile modeling in earlier sections.

Note that a more realistic treatment of the problem implies there would be some variation of the polarization at redshifts as well as at blueshifts.  Indeed, observations of such an effect would help constrain refinements of both the model geometry of the CIR as well as the distribution of scattered light throughout the optically thick wind, a topic for a future study.

\bsp    % typesetting comment
\label{lastpage}

\end{document}